\documentclass[a4paper, aps, prl, superscriptaddress, 10pt]{revtex4-2}

\usepackage{amsfonts}
\usepackage{amssymb}
\usepackage{amsmath}
\usepackage[utf8]{inputenc}
\usepackage[T1]{fontenc}
\usepackage[english]{babel} 
\usepackage{graphicx}
\usepackage{hyperref}
\hypersetup{colorlinks, citecolor=black, linkcolor=black, urlcolor=black}
\usepackage{comment}
\usepackage{siunitx}
\usepackage{float}
\usepackage{placeins}
\usepackage{subcaption}
\usepackage[font=small,labelfont=bf]{caption}
\usepackage{textcomp, gensymb} 
\usepackage[version=4]{mhchem}
\usepackage{multirow}
 
\usepackage{textgreek}

\usepackage{silence}
\begin{document}
\title{Direct detection of hydrogen reveals a new macroscopic crustal water reservoir on early Mars}

\author{Estrid Buhl Naver}
\thanks{These authors contributed equally.}
\affiliation{Department of Energy Conversion and Storage, Technical University of Denmark, Kgs. Lyngby, Denmark}
\affiliation{Department of Physics, Technical University of Denmark, Kgs. Lyngby, Denmark}

\author{Katrine Wulff Nikolajsen}
\thanks{These authors contributed equally.}
\affiliation{Centre for Star and Planet Formation, Globe Institute, University of Copenhagen, Copenhagen, Denmark}

\author{Martin Sæbye Carøe}
\affiliation{Department of Applied Mathematics and Computer Science, Technical University of Denmark, Kgs. Lyngby, Denmark}

\author{Domenico Battaglia}
\affiliation{Department of Energy Conversion and Storage, Technical University of Denmark, Kgs. Lyngby, Denmark}

\author{Jens Frydenvang}
\affiliation{Centre for Star and Planet Formation, Globe Institute, University of Copenhagen, Copenhagen, Denmark}

\author{Martin Bizzarro}
\affiliation{Centre for Star and Planet Formation, Globe Institute, University of Copenhagen, Copenhagen, Denmark}

\author{Jakob Sauer Jørgensen}
\affiliation{Department of Applied Mathematics and Computer Science, Technical University of Denmark, Kgs. Lyngby, Denmark}

\author{Kim Lefmann}
\affiliation{Nanoscience Center, Niels Bohr Institute, University of Copenhagen, Copenhagen, Denmark}

\author{Anders Kaestner}
\affiliation{PSI Center for Neutron and Muon Sciences, Paul Scherrer Institut, Villigen, Switzerland}

\author{David Christian Mannes}
\affiliation{PSI Center for Neutron and Muon Sciences, Paul Scherrer Institut, Villigen, Switzerland}

\author{Phil Cook}
\affiliation{European Synchrotron Radiation Facility, Grenoble, France}
\affiliation{Danish Technological Institute, Høje Taastrup, Denmark}

\author{Henrik Birkedal}
\affiliation{Department of Chemistry, Aarhus University, Aarhus, Denmark}

\author{Thorbjørn Erik Køppen Christensen}
\affiliation{Department of Applied Mathematics and Computer Science, Technical University of Denmark, Kgs. Lyngby, Denmark}
\affiliation{MAX IV Laboratory, Lund University, Lund, Sweden}

\author{Innokenty Kantor}
\affiliation{Department of Physics, Technical University of Denmark, Kgs. Lyngby, Denmark}
\affiliation{MAX IV Laboratory, Lund University, Lund, Sweden}

\author{Henning Friis Poulsen}
\affiliation{Department of Physics, Technical University of Denmark, Kgs. Lyngby, Denmark}

\author{Luise Theil Kuhn}
\affiliation{Department of Energy Conversion and Storage, Technical University of Denmark, Kgs. Lyngby, Denmark}

\date{\today}

\begin{abstract}
The next great leap in Martian exploration will be the return of samples to Earth. To ensure the maximum scientific return from studying these samples, the development and utilisation of non-destructive analytical techniques are essential to enable early three-dimensional characterisation of their interiors. Neutron computed tomography is a powerful method in this context: it is highly sensitive to hydrogen and complements the more conventional X-ray computed tomography. Because the distribution and nature of hydrous phases are central to understanding the habitability, the climatic and geological evolution, and potential biosignatures of Mars, identifying hydrogen-bearing phases in Martian crustal rocks is of particular importance. Using the only Martian crustal material available on Earth, the NWA 7034 meteorite and its pairs, we show that combined neutron and X-ray computed tomography enables non-destructive sample-wide mapping of hydrogen and reveals the distribution and petrographic contexts of hydrous phases. We identify hydrogen-rich iron oxyhydroxides within ancient igneous clasts, forming a macroscopic mineralogical water reservoir within the meteorite. These alteration assemblages closely resemble those observed in samples collected by the Perseverance rover in Jezero crater, where hydrated iron oxyhydroxides are also present. This similarity suggests that such phases may represent a widespread near-surface water reservoir on early Mars.

\end{abstract}

\maketitle

\section{Introduction}

With the successful return of extraterrestrial material from the Itokawa asteroid by Hayabusa \cite{Fujiwara2006}, the Ryugu asteroid by Hayabusa2 \cite{Watanabe2017}, the Bennu asteroid by OSIRIS-REx \cite{Lauretta2017}, and the ongoing Mars Sample Return (MSR) efforts \cite{Muirhead2020}, we have entered a new era of sample-return missions. The Perseverance rover is collecting a carefully curated assemblage of samples in Jezero crater to be returned to Earth as part of the MSR campaign \cite{Farley2020}. This collection comprises diverse geological specimens, including igneous, sedimentary, water-altered, and impact-related rocks, as well as one sample containing a potential biosignature \cite{Herd2025, Hurowitz2025, Simon2023}. Their return to specialised laboratories on Earth will provide crucial new insights into the environmental, geological, and hydrological evolution – and potentially the prebiotic chemistry and biology – of Mars. For this reason, the completion of MSR is considered the highest scientific priority of robotic exploration efforts this decade \cite{NationalAcademies2022}. Developing analytical approaches that maximise the scientific return of studying these irreplaceable samples is therefore critical.

Non-destructive methods will play a key role in the early stages of sample characterisations, as they provide essential knowledge of the sample's interior before it is subjected to destructive analyses. The conventional approach to characterising geological samples relies heavily on destructive analyses of polished sections cut from the exterior, which has two significant limitations. First, many high-precision characterisation techniques (such as chemical and isotope composition analyses) require that samples are crushed and/or dissolved. Second, sections are often prepared without prior knowledge of internal heterogeneity or three-dimensional structure. Integrating non-destructive techniques can help address these issues by enabling interior characterisation in three dimensions while preserving the sample’s pristineness, which provides valuable guidance for following analyses, prevents unnecessary sample loss, and preserves a record of the interior structures of a sample. Presently, X-ray computed tomography (XCT) is the primary suggested technique for early, non-destructive characterisation of the MSR samples \cite{Tait2022}, but including complementary non-destructive methods is highly advantageous. 

Neutron computed tomography (NCT) is such a complementary non-destructive approach for three-dimensional imaging \cite{Oestergaard2023, Battaglia2025} that is generally underutilised for geological materials. Unlike the more commonly used XCT, where attenuation is proportional to atomic number and material density, neutron attenuation depends more erratically on atomic number. As a result, NCT and XCT offer different sensitivities to elements and isotopes, making them highly complementary. Aside from its non-destructive nature, NCT is especially promising for analysing returned samples because neutrons can penetrate higher-Z materials (such as titanium tubes enclosing samples collected by the Perseverance rover) more easily than X-rays, it requires no sample preparation, and it is highly sensitive to hydrogen \cite{Fedrigo2018, Martell2024}. Because hydrogen is a key tracer of water and potential habitability, this sensitivity makes NCT particularly suited for analysing planetary samples, by offering a direct detection of hydrous and potentially organic material at a whole-sample scale. The individual mineral phases hosting the H can then be identified by X-ray diffraction. This can also be done tomographically by X-ray diffraction computed tomography (XRD-CT) \cite{Bleuet2008, Stock2008, Birkbak2015, Frølich2016, Christensen2024}.

Understanding the past and present distribution of water, and how water has interacted with the rock record on other planetary bodies, is fundamental to assessing habitability and searching for biosignatures \cite{MEPAG2020, NationalAcademies2022}.  Although liquid water is not stable on the Martian surface today, it exists as H$_2$O ice at the poles \cite{Boynton2002, Feldman2002, Levrard2004, Mellon2004}, bound to minerals in the regolith and exposed bedrock \cite{Feldman2002, Wernicke2021, Audouard2014, Milliken2007, Bibring2006, Carter2013}, and possibly incorporated in dust \cite{Valantinas2025, Chen2021}. Orbital mineral mapping reveals a temporal shift in the dominant alteration mineral assemblages, consistent with a transition from an early warm and wet environment to the present cold and arid conditions \cite{Bibring2006}. This supports the hypothesis that surface liquid water persisted for extended periods in early Martian history. In-situ rover observations further corroborate this: the Curiosity rover has documented fluvial and lacustrine deposits in Gale crater, indicating millions of years of aqueous activity  \cite{Vasavada2022, Grotzinger2015}, and the recent discovery of well-defined wave ripples \cite{Mondro2025} extends this record of surface liquid water even further.

Here, we investigate the potential of combining NCT, XCT, and XRD-CT to map hydrogen distribution in samples from Mars. As a representative material, we analyse the Northwest Africa (NWA) 7034 meteorite. NWA 7034 and its paired stones are the only meteorites that directly sample the ancient surface crust of Mars. Derived from the southern highlands \cite{Lagain2022, Herd2024}, it is a polymict impact breccia containing diverse lithic clasts, including basaltic and evolved igneous, sedimentary, and impact melt clasts, embedded in a fine-grained matrix \cite{Agee2013, Wittmann2015, Santos2015}. It also hosts the oldest known igneous material from Mars, with zircon crystallization ages up to 4.48 Ga \cite{Bouvier2018, Humayun2013, Mccubbin2016, Hu2019} and is notably water-rich compared to other Martian meteorites, with bulk concentrations of 6,000 ppm H$_2$O \cite{Agee2013}. These characteristics make it the best available analogue for the samples collected in Jezero crater by the Perseverance rover. Whereas previous studies have mainly inferred hydrogen contents from the stoichiometry of hydrous minerals in the meteorite matrix \cite{Muttik2014}, we present methods that detect hydrogen directly and non-destructively. This approach enables whole-sample mapping of hydrogen, allowing identification of its mineral hosts and petrographic contexts, and offers new insights into surficial water–rock interactions on early Mars. 

\section{Results}
Neutron and X-ray tomograms of a 12x8x2 mm$^3$ piece of NWA 7034 (Fig. 1a) reveal a complex internal microstructure comprising diverse monomineralic and polymineralic clasts embedded in a fine-grained matrix. Combined NCT-XCT enable classifying the sample volume into four likely categories of minerals: 1) feldspars, characterised by low neutron and X-ray attenuation; 2) Fe-Ti oxides, showing high X-ray and medium neutron attenuation; 3) H-rich Fe-oxyhydroxides, exhibiting high attenuation in both modalities; and 4) the matrix, consisting of fine-grained material (<1 voxel/6.2 \textmu m$^3$), including pyroxene, plagioclase, phosphate, and Fe-Ti oxides;.

Phase segmentation of the putative Fe-Ti oxides and H-rich Fe-oxyhydroxides (Fig. 1b) show that Fe-Ti oxides occur predominantly as small grains dispersed throughout the meteorite and comprising 3 vol\% of the sample. The H-rich phases form localised macroscopic clusters totalling 0.4 vol\%. The three largest H-rich regions (Fig. 1c) were selected for detailed analysis to investigate the nature of the H-rich phases. They co-appear with Fe-Ti oxides (Fig. 1c) and are contained within distinct lithic clasts ranging from 120x270x150 \textmu m$^3$ to 300x380x330 \textmu m$^3$ in size (Fig. 2), hereafter termed H-Fe-ox clasts. These clasts host pronounced hydrogen hotspots in NCT and correspondingly high X-ray attenuation (Fig. 2).

\begin{figure}[h]
    \centering
    \includegraphics[width=0.9\linewidth]{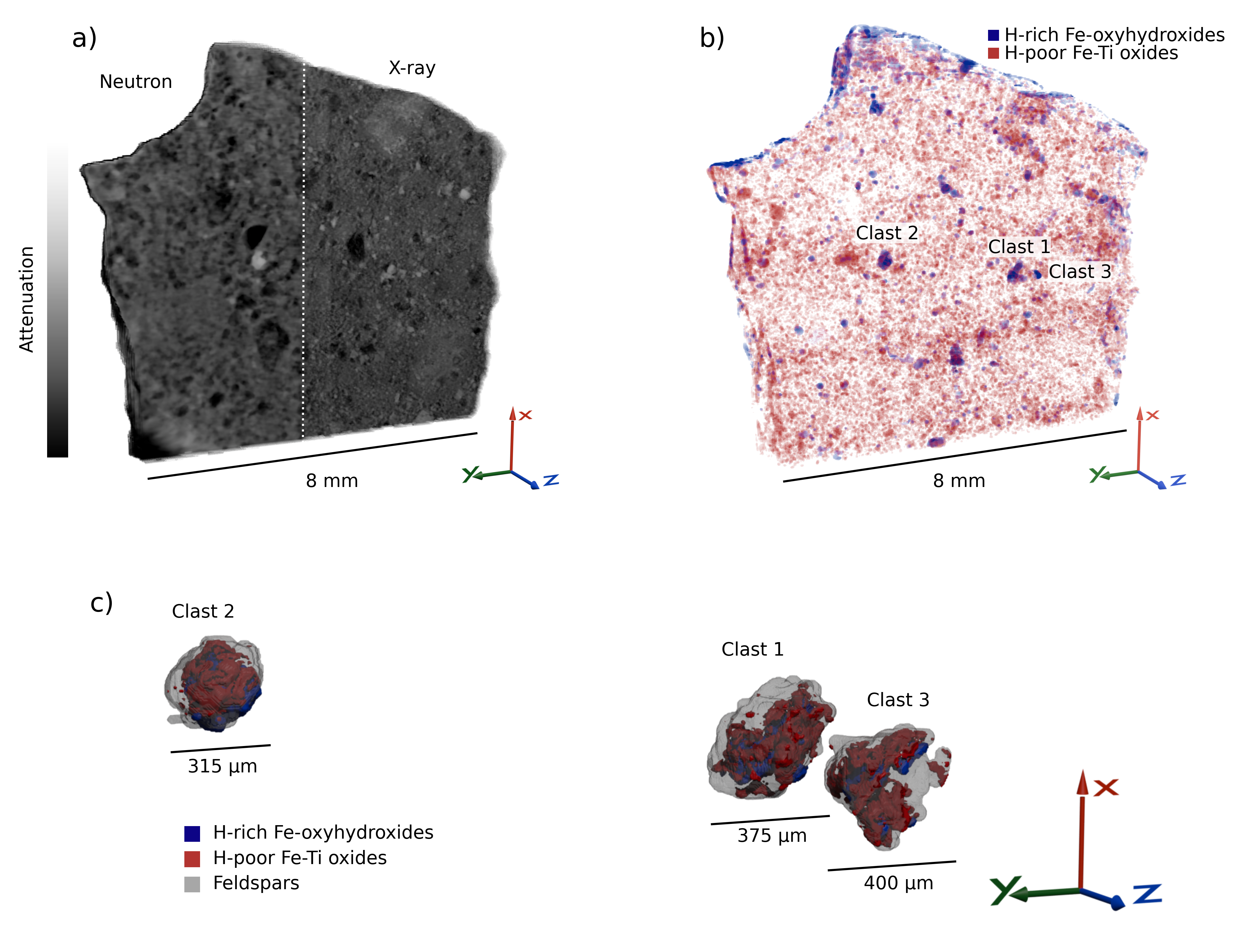}
    \caption{\textbf{3D rendering of meteorite and clasts.} a) 3D rendering of the attenuation tomograms cut in half in $z$ to show two H-Fe-ox clasts, showing neutron attenuation on the left and X-ray attenuation on the right. White means high attenuation, dark means low attenuation. b) 3D rendering of segmented phases throughout the meteorite, with H-rich Fe-oxyhydroxides in blue and H-poor Fe-Ti oxides in red. c) 3D rendering of the three biggest H-Fe-ox clasts.} 
    \label{Fig. seg}
\end{figure}

All three H-Fe-ox clasts consist of three distinct petrographic components (Fig. 2): (1) medium neutron attenuating Fe-Ti oxides, including an euhedral grain at 170x180x170 \textmu m$^3$; (2) high neutron attenuating Fe-oxyhydroxides, which do not show clearly defined single grains but cover volumes up to 120x350x230 \textmu m$^3$; and (3) low X-ray and neutron attenuating feldspars forming subhedral interlocking grains up to 100x440x150 \textmu m$^3$ that surround the Fe-Ti oxides. XRD-CT shows that the clasts also display a small-angle X-ray scattering (SAXS) signal (diffracted signal at scattering vector 0.1 Å$^{-1}$<Q<0.6 Å$^{-1}$) coincident with the most neutron attenuating regions. A representable matrix area (Fig. 2) shows minor Fe-Ti-oxides but no H hotspots or SAXS signal.

\begin{figure}[h]
    \centering
    \includegraphics[width=0.95\linewidth]{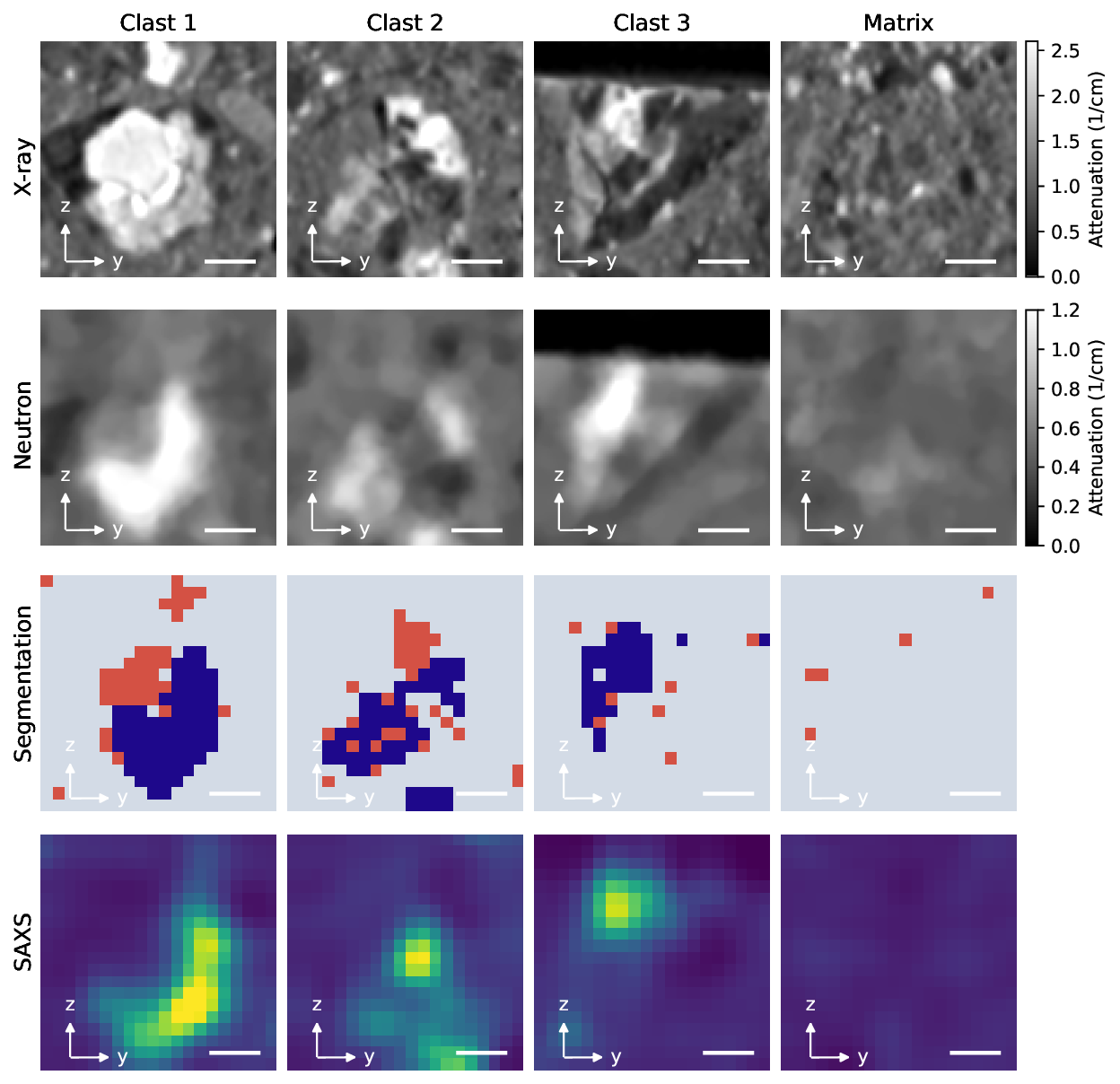}
    \caption{\textbf{Close up view of H-Fe-Ox clasts and matrix.} The rows show different contrasts and the columns show different areas in the meteorite. The contrasts are X-ray (XCT) and neutron (NCT) imaging in row 1 and 2, respectively, as well as the H-rich (blue) and Fe/Ti, H-poor (red) segmented phases from XCT and NCT downscaled to match XRD-CT resolution in row 3 and the SAXS (XRD-CT) signal (diffracted signal at scattering vector 0.1 Å$^{-1}$<Q<0.6 Å$^{-1}$) in row 4. All scale bars are 100 \textmu m.}
    \label{Fig. ROI_clasts}
\end{figure}

To identify the mineral hosts associated with the H-rich regions, Rietveld refinement is performed on the average X-ray diffraction from the low neutron attenuating (H-poor) Fe-Ti oxides and high neutron attenuating (H-rich) Fe-oxyhydroxides segmented pixels in H-Fe-ox clasts. The dominant identified phases are plagioclase feldspar, ilmenite, and magnetite/maghemite, together with minor amounts of rutile (Fig. 3). The diffraction patterns suggest trace amounts of 1-2 other additional phases (Fig. S1 and S2), which cannot be identified either because they appear in too small quantities or are too weakly scattering. Common for all three clasts is the high amount of magnetite/maghemite in the H-rich areas, along with a smaller amount of ilmenite and rutile.

\begin{figure}[h]
    \centering
    \includegraphics[width=0.5\linewidth]{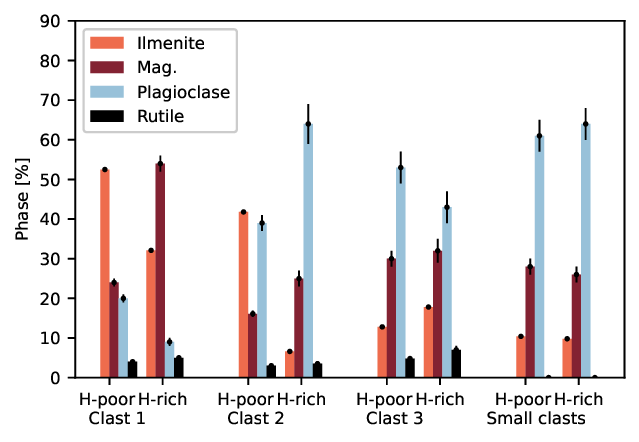}
    \caption{ \textbf{Composition of H-Fe-ox clasts.} Comparison between compositions of H-rich and H-poor areas in the H-Fe-Ox clasts based on Rietveld refinement of XRD-CT data.  "Mag." represents magnetite/maghemite.}
    \label{Fig. phases}
\end{figure}

To quantify hydrogen content in the H-Fe-ox clasts and matrix, we compare measured X-ray and neutron attenuation with theoretical values calculated from idealised mineral compositions (Section S2). For the matrix, theoretical attenuations of 0.96 cm$^{-1}$ (X-ray) 0.37 cm$^{-1}$  (neutron), based on mineralogy from \cite{Agee2013}, is identical to the measured X-ray attenuation but slightly lower than the measured neutron attenuation (0.45 cm$^{-1}$). For the H-Fe-ox clasts, regions with low H signal exhibit close agreement between theoretical (1.7 cm$^{-1}$ X-ray; 0.56 cm$^{-1}$ neutron) and measured attenuations (1.8 cm$^{-1}$; 0.57 cm$^{-1}$). In the H-rich regions, however, the measured attenuations (3.4 cm$^{-1}$ X-ray; 1.9 cm$^{-1}$ neutron) exceed theoretical values (2.0 cm$^{-1}$; 0.61 cm$^{-1}$). 

We assume that all excess measured neutron attenuation results from the presence of hydrogen. To estimate the H concentration, it is necessary to know in which configuration hydrogen exists, e.g., as H, OH, or \ce{H2O}, which depends on the nature of the hydrated mineral phases and their formation paths. Infrared spectroscopy measurements of NWA 7034 \cite{Beck2015} indicate that H is primarily incorporated as structural OH. Adopting this assumption, we estimate that H-Fe-ox clasts contain up to 15 wt.\% OH. Combined with their volumetric abundance (0.4 vol\%), these clasts contribute 635 ppm \ce{H2O} or 11 wt.\% of the meteorite’s 6,000 ppm bulk water content. 
A similar estimation is made for the matrix and results in 0.6 wt.\% OH, corresponding to 6,000 ppm. 

The excess measured X-ray attenuation is assumed to be from other Fe-ox phases with diffraction signal too weak to show up in the XRD-CT.

\section{Discussion}
This three-dimensional mapping of hydrogen in a 12x8x2 mm$^3$ section of NWA 7034 reveals pronounced and highly localised hydrated areas within the meteorite. While previous studies of NWA 7034 have inferred hydration indirectly from thin-section analyses, our combined NCT-XCT-XRDCT approach provides the first sample-wide, direct, and non-destructive mapping of hydrogen in a Martian regolith sample. Our studies reveal that Fe-oxyhydroxides within igneous clasts represent a macroscopic reservoir of concentrated water in the meteorite. At up to 635 ppm \ce{H2O}, these clasts account for 11\% of the total water content of NWA 7034.

Contrary to the matrix, which represents a mix of fine grains and larger fragments with unknown origins, ages, alteration histories, and internal genetic relationships, the igneous clasts all originate from distinct source rocks where these origins are ascertainable. Moreover, it is possible that this type of igneous clasts is the main source of hydrated Fe-Ti-oxyhydroxides found in the matrix, appearing as loose fragments broken off from the larger clasts. Their mineralogy and textures match the Fe-Ti-rich plutonic clasts described in previous work \cite{Santos2015, Agee2013, Wittmann2015, Hu2019, Hewins2017}. These contain Fe-Ti oxide and feldspar as common phases along with apatite and, more rarely, pyroxene, which the H-Fe-ox clasts lack. H-Fe-ox clasts are most similar to Fe, Ti, and P (FTP) clasts \cite{Santos2015} based on the presence of large and subhedral to euhedral grains of Fe-Ti oxide. While FTP clasts contain apatites with metamorphic ages of 1,500 Ma \cite{Mccubbin2016}, Pb-Pb dating of zircons in igneous lithologies in NWA 7034 reveal crystallisation ages between 4,450 and 4,200 Ma \cite{Hu2019, Humayun2013, Mccubbin2016, Bouvier2018, Costa2020}. Although zircons have not yet been found in FTP clasts, they are likely to have similar crystallization ages to other igneous clasts \cite{Nyquist2016}.


The distribution of H in the H-Fe-ox clasts is spatially correlated with magnetite/maghemite and, to a lesser extent, ilmenite, as seen in Fig. 3. Based on the XRD-CT data, however, it is not possible to determine which of these phases (if any) is the primary host of H. Magnetite and ilmenite are common magmatic minerals widespread in igneous rocks and in NWA 7034, whereas maghemite is an alteration product formed by oxidation of magnetite or by heating of other Fe-oxides such a ferrihydrite \cite{Cornell2003}. All phases identified in the H-Fe-ox clasts are nominally anhydrous, implying that they must either be hydrated or associated with hydrated components not resolved in the XRD-CT data. Previous work has attributed water hosted by matrix Fe-oxides to maghemite \cite{Muttik2014}, goethite \cite{Gattacceca2014}, and possibly ferrihydrite \cite{Muttik2014}. Multiple observations indicate that primary Fe-Ti oxides in NWA 7034 have undergone alteration: magnetite shows partial transformation to maghemite \cite{Gattacceca2014}, while the presence of rutile in the H-Fe-ox clasts indicates ilmenite alteration \cite{Mucke1991}. Such alteration processes can introduce OH, protons, or \ce{H2O} onto Fe-oxide surfaces or in crystal lattices, as e.g., in maghemite which can incorporate up to 2 wt\% \ce{H2O} \cite{Cornell2003}.

Several lines of evidence indicate that additional hydrated phases must be present in H-Fe-ox clasts. First, although H-regions contain 20-50 wt\% maghemite, this contribution alone can only account for a minor amount of the total estimated water content in the clasts. Second, XRD-CT patterns from H-rich phases reveal signals from trace amounts of one or two additional phases (Fig. S4,5), which cannot be confidently identified due to their low abundance or weak diffraction signal. Third, the H-rich areas exhibit excess X-ray attenuation relative to theoretical values based on the detected mineralogy, suggesting the presence of additional highly X-ray attenuating phases. Fourth, the strong SAXS signal characteristic of 5 nm structures occurs exclusively within the H-rich regions. Collectively, these features imply that the dominant H host is one or more nanophase alteration phases, plausibly including goethite or ferrihydrite. 


Water released by stepwise heating of the meteorite has $\Delta^{17}$O values that plot above the Terrestrial Fractionation Line, indicating that the material is predominantly, if not completely, Martian in origin \cite{Agee2013}. Small amounts of Fe oxyhydroxides that partially replace pyrite grains have been attributed to terrestrial alteration \cite{Lorand2015}. There is, however, no indication that the H-Fe-ox clasts formed on Earth, since this would be evident in the O-isotopic composition of water released at low temperatures, considering their relatively large contribution to the overall water budget. 

Similar to igneous clasts in NWA 7034, both lava flows of the Máaz formation, and the olivine-bearing cumulates of the Séítah formation contain primary Fe-Ti-oxides \cite{Farley2022, Wiens2022, Mandon2023, Schmidt2025, Tice2022}. Moreover, Fe-oxyhydroxides contribute to the secondary mineral assemblages \cite{Farley2022, Wiens2022, Mandon2023}, and particularly in the Máaz formation, a mixture of these and phyllosilicates is likely the main carriers of hydration \cite{Mandon2023}. The close similarity between alteration phases in NWA 7034 and those identified by Perseverance in Jezero crater suggests that the hydrothermal or weathering processes recorded in NWA 7034 were widespread on early Mars. This is recorded in the collected samples where the pairs Salette/Coulette and Robine/Malay collected from the Séítah formation, and Hahonih/Astah from the Máaz formation, all contain secondary Fe-oxyhydroxides \cite{Simon2023}. These parallels underscore the value of NWA 7034 as an analogue for the MSR collection. Our results demonstrate how combined NCT-XCT-XRDCT can enhance early-stage characterisation of returned samples by directly mapping hydrogen in three dimensions, offering a preview of the unique insights these methods will provide into the distribution, nature, and context of hydrous phases in MSR samples.

\section{Methods}
\subsection{Neutron and X-ray tomography}
A 12x8x2 mm$^3$ piece of NWA 7034 was used for characterisation. For all experiments, it was glued to a stick, then placed on a rotation stage. One surface was polished by hand after the XCT and before the NCT and XRD-CT for characterisation unrelated to the current study.

X-ray imaging was performed at the BM05 beamline at the European Synchrotron Radiation Facility in France using a voxel size of 3 \textmu m$^3$, 6000 projections with exposure time 50 ms, and a white beam with mean energy 73 keV. In order to image the full sample, 7 tomograms were acquired with different vertical offsets.

Neutron imaging was performed at the ICON beamline \cite{Kaestner2011} at the Paul Scherrer Institute in Switzerland, using a white beam with peak energy 25 meV. The sample was measured using the taper setup \cite{Morgano2018} with a circular field-of-view of diameter 10 mm, a voxel size of 6.2 \textmu m$^3$, and an exposure time of 4 min per projection. To limit noise, each projection was calculated as the median of four times 60 s measurements. The neutron beam was collimated with a circular aperture of diameter D=20 mm, L=7 m from the sample position, resulting in L/D = 350. The sample was rotated 360\degree\ and 1125 projections were measured.

All projections were flat and dark field corrected. Neutron projections were noise filtered to remove gamma ray spots, using the spot removal filter from MuhRec \cite{Kaestner2011_2}. X-ray projections were filtered using Paganin phase retrieval \cite{Paganin2002} to extract attenuation contrast. Filter strengths were tuned empirically.
For tomographic reconstruction, we used the Core Imaging Library \cite{Jorgensen2021, Pasca2025} and the ASTRA toolbox \cite{vanAarle2016}. X-ray data was reconstructed with Filtered Back-Projection. We used an optimization-based iterative reconstruction method for the neutron data with total variation regularization. The regularization parameter for the total variation term was chosen empirically to balance noise suppression and preservation of structural details. The optimization was carried out using the FISTA algorithm \cite{Beck2009, Beck2009_2}, and iterations were stopped once the change in the objective function became negligible, indicating convergence.. The 7 X-ray tomograms were stitched vertically into a single volume, then registered to the neutron reconstructions using SimpleITK \cite{Lowekamp2013}, yielding a 3D volume with 2 intensities at each spatial location in the sample. Segmentation was performed using 2D thresholding to separate into four phases: low attenuation, matrix, highly X-ray attenuating with low neutron attenuation and highly neutron attenuating.

\subsection{X-ray diffraction tomography}
X-ray diffraction tomography \cite{Bleuet2008, Stock2008, Birkbak2015, Frølich2016, Christensen2024, Vamvakeros2018} was performed at the DanMAX beamline at the MAX IV synchrotron in Sweden. Diffraction patterns were recorded using a 35 keV beam with a size of 28x28 \textmu m$^2$ scanning across the sample in 25 \textmu m steps, then rotating the sample 1\degree, scanning again, and so on. Measurements were recorded at a frequency of 200 Hz using an Pilatus CdTe 2M detector to record the scattering in 2D. The distance between sample and detector was 600 mm. 
This produces a dataset similar to that of an attenuation tomography, but instead of attenuation projections there are diffraction projections where each pixel contains a diffraction pattern for that location and sample rotation.

The 2D diffraction patterns were azimuthally integrated using the MatFRAIA algorithm \cite{Jensen2022} and reconstructed using the gridrec algorithm to get a 3D volume with a diffraction pattern in each voxel including small angle X-ray scattering. Diffraction patterns from the segmented voxels in clasts 1,2, and 3 were averaged together and Rietveld refined using Maud \cite{Lutterotti2000} and crystal structures for plagioclase feldspar \cite{Curetti2011}, ilmenite \cite{Morosin1978}, magnetite \cite{Wechsler1984}, and rutile \cite{Ballirano2001}.

\subsection{Attenuation calculations}
Mineral attenuation values for neutrons and X-rays were calculated from weighted means of tabulated element values, specifically mass X-ray attenuations \cite{NIST_xray} and total neutron cross section \cite{NIST_neutron}, and tabulated mineral densities \cite{Ralph2025} (Eq. S1). The weights were found using the chemical compositions of the minerals. The different minerals are considered as phases of fixed composition rather than solid solutions, e.g. albite and anorthite represent idealised compositions of the two types of plagioclase feldspar. The plagioclase attenuation is given as a mean of the two types of plagioclase. Clinopyroxene is also given as an average between two versions. A list of the compositions can be found in Table S1. The isotope composition of the elements is taken to be the same as for elements on Earth, as the potential difference in composition of e.g. H and O does not result in significant differences in the calculated neutron attenuation.

When estimating OH contents in the clasts and matrix, the mass attenuation is used which is multiplied with the density of the meteorite to find the linear attenuation (Eq. S2).

\section{Acknowledgements}
The work of E.B.N was supported by the ESS Lighthouse on Hard Materials in 3D, SOLID, funded by the Danish Agency for Science and Higher Education (8144-00002B) and the Danish Agency for Science, Technology, and Innovation for funding the instrument center DanScatt. The work of K.W.N was supported by ESA and the Villum foundation.
The work of M.S.C. was supported by Danish Data Science Academy, funded by the Novo Nordisk Foundation (NNF21SA0069429) and the Villum foundation (40516). This work made use of computational support by CoSeC, the Computational Science Centre for Research Communities, through CCPi (EPSRC grant EP/T026677/1).
This work is based on experiments P20230605 conducted at the Swiss Spallation Neutron Source SINQ, Paul Scherrer Institute, Villigen, Switzerland, 20240396 conducted at MAX-IV, Lund, Sweden, and the European Synchrotron Radiation Facility, Grenoble, France. Research conducted at MAX IV, a Swedish national user facility, is supported by Vetenskapsrådet (Swedish Research Council, VR) under contract 2018-07152, Vinnova (Swedish Governmental Agency for Innovation Systems) under contract 2018-04969 and Formas under contract 2019-02496.

\section{Data availability}
Data will be uploaded to data.dtu.dk.

\section{Author contributions}
Conceptualisation: E.B.N., K.W.N, J.F., K.L., L.T.K., H.F.P, M.B., H.B.
Investigation: E.B.N., K.W.N, D.B., A.K., D.C.M., P.C., T.E.K.C., I.K., H.F.P., L.T.K., H.B.
Software: E.B.N, M.S.C., D.B., T.E.K.C. 
Formal Analysis: M.S.C., J.S.J.
Supervision: J.F., K.L., L.T.K., H.F.P, M.B., J.S.J., H.B.
Writing – original draft: E.B.N, K.W.N, M.S.C
Writing - review and editing: Everyone.

\bibliography{bibliography} 

@article{Agee2013,
    author = {C. B. Agee and others},
    title = {{Unique Meteorite from Early Amazonian Mars: Water-Rich Basaltic Breccia Northwest Africa 7034}},
    journal = {Science},
    volume = {339},
    number = {6121},
    pages = {780-785},
    year = {2013},
    doi = {10.1126/science.1228858},
    URL = {https://www.science.org/doi/abs/10.1126/science.1228858},
}

@article{Audouard2014,
    author = {Audouard, Joachim and others}, 
    title = {{Water in the Martian regolith from OMEGA/Mars Express}},
    journal = {Journal of Geophysical Research: Planets},
    volume = {119},
    number = {8},
    pages = {1969-1989},
    year = {2014},
    doi = {10.1002/2014JE004649},
    URL = {https://agupubs.onlinelibrary.wiley.com/doi/full/10.1002/2014JE004649},
}

@article{Ballirano2001,
    author = "Ballirano, Paolo and Caminiti, Ruggero",
    title = "{Rietveld refinements on laboratory energy dispersive X-ray diffraction (EDXD) data}",
    journal = "Journal of Applied Crystallography",
    year = "2001",
    volume = "34",
    number = "6",
    pages = "757--762",
    month = "Dec",
    doi = {10.1107/S0021889801014728},
    url = {https://doi.org/10.1107/S0021889801014728},
}

@article{MEPAG2020,
  title={{MEPAG (2020), Mars Scientific Goals, Objectives, Investigations, and Priorities}},
  author={Banfield, D and others},
  journal={White paper posted March, 2020 by the Mars exploration program analysis group (MEPAG)},
  pages={89},
  year={2020}
}

@article{Battaglia2025,
    title = {{Neutron imaging in 2D and 3D as a powerful tool to investigate electrolyte degradation and plating mechanisms in sodium-ion batteries}},
    journal = {Journal of Power Sources},
    volume = {655},
    pages = {237846},
    year = {2025},
    issn = {0378-7753},
    doi = {https://doi.org/10.1016/j.jpowsour.2025.237846},
    url = {https://www.sciencedirect.com/science/article/pii/S0378775325016829},
    author = {Domenico Battaglia and Estrid Buhl Naver and Cédric Holme Qvistgaard and Matteo Busi and Pavel Trtik and Markus Strobl and Anna Fedrigo and Annabel Olgo and Alar Jänes and Nikolaj Zangenberg and Søren Schmidt and Luise Theil Kuhn},
}

@ARTICLE{Beck2009,
  author={Beck, Amir and Teboulle, Marc},
  journal={IEEE Transactions on Image Processing}, 
  title={{Fast Gradient-Based Algorithms for Constrained Total Variation Image Denoising and Deblurring Problems}}, 
  year={2009},
  volume={18},
  number={11},
  pages={2419-2434},
  doi={10.1109/TIP.2009.2028250}
}

@article{Beck2009_2,
    author = {Beck, Amir and Teboulle, Marc},
    title = {{A Fast Iterative Shrinkage-Thresholding Algorithm for Linear Inverse Problems}},
    journal = {SIAM Journal on Imaging Sciences},
    volume = {2},
    number = {1},
    pages = {183-202},
    year = {2009},
    doi = {10.1137/080716542}, 
    URL = { https://doi.org/10.1137/080716542},
}

@article{Beck2015,
    title = {{A Noachian source region for the “Black Beauty” meteorite, and a source lithology for Mars surface hydrated dust?}},
    journal = {Earth and Planetary Science Letters},
    volume = {427},
    pages = {104-111},
    year = {2015},
    issn = {0012-821X},
    doi = {10.1016/j.epsl.2015.06.033},
    url = {https://www.sciencedirect.com/science/article/pii/S0012821X15003969},
    author = {P. Beck and others}
}

@Article{Bleuet2008,
    author ="Bleuet, Pierre and Welcomme, Eléonore and Dooryhée, Eric and Susini, Jean and Hodeau, Jean-Louis and Walter, Philippe",
    title  ="{Probing the structure of heterogeneous diluted materials by diffraction tomography}",
    journal ="Nature Materials",
    year  ="2008",
    volume  ="7",
    issue  ="6",
    pages  ="468-472",
    doi  ="10.1038/nmat2168",
    url  ="https://doi.org/10.1038/nmat2168",
}

@article{Bibring2006,
    author = {Jean-Pierre Bibring and others},
    title = {{Global Mineralogical and Aqueous Mars History Derived from OMEGA/Mars Express Data}},
    journal = {Science},
    volume = {312},
    number = {5772},
    pages = {400-404},
    year = {2006},
    doi = {10.1126/science.1122659},
    URL = {https://www.science.org/doi/abs/10.1126/science.1122659},
}

@Article{Birkbak2015,
    author ="Birkbak, M. E. and Leemreize, H. and Frølich, S. and Stock, S. R. and Birkedal, H.",
    title  = "{Diffraction scattering computed tomography: a window into the structures of complex nanomaterials}",
    journal  ="Nanoscale",
    year  ="2015",
    volume  ="7",
    issue  ="44",
    pages  ="18402-18410",
    publisher  ="The Royal Society of Chemistry",
    doi  ="10.1039/C5NR04385A",
    url  ="http://dx.doi.org/10.1039/C5NR04385A",
}

@article{Bouvier2018,
    author = {L. C. Bouvier and others},
    doi = {10.1038/s41586-018-0222-z},
    journal = {Nature},
    pages = {586-589},
    title = {{Evidence for extremely rapid magma ocean crystallization and crust formation on Mars}},
    volume = {558},
    url = {https://doi.org/10.1038/s41586-018-0222-z},
    year = {2018} 
}

@article{Boynton2002,
    title = {{Distribution of Hydrogen in the Near Surface of Mars: Evidence for Subsurface Ice Deposits}},
    author={W. V. Boynton and others}, 
    journal={Science},
    volume={297},
    number={5578},
    pages={81--85},
    year={2002},
    doi = {10.1126/science.1073722},
    URL = {https://www.science.org/doi/abs/10.1126/science.1073722},
}

@article{Carter2013,
  title={{Hydrous minerals on Mars as seen by the CRISM and OMEGA imaging spectrometers: Updated global view}},
  author={J. Carter and F. Poulet and J.-P. Bibring and N. Mangold and S. Murchie},
  journal={Journal of Geophysical Research: Planets},
  volume={118},
  number={4},
  pages={831--858},
  year={2013},
  publisher={Wiley Online Library}
}

@article{Chen2021,
    author = {Chen, Si Athena and Heaney, Peter J. and Post, Jeffrey E. and Fischer, Timothy B. and Eng, Peter J. and Stubbs, Joanne E.},
    title = {{Superhydrous hematite and goethite: A potential water reservoir in the red dust of Mars?}},
    journal = {Geology},
    volume = {49},
    number = {11},
    pages = {1343-1347},
    year = {2021},
    issn = {0091-7613},
    doi = {10.1130/G48929.1},
    url = {https://doi.org/10.1130/G48929.1},
}

@article{Christensen2024,
    title = {{3D distribution of biomineral and chitin matrix in the stomatopod dactyl club by high energy XRD-CT}},
    journal = {Journal of Structural Biology},
    volume = {216},
    number = {4},
    pages = {108136},
    year = {2024},
    issn = {1047-8477},
    doi = {10.1016/j.jsb.2024.108136},
    url = {https://www.sciencedirect.com/science/article/pii/S1047847724000765},
    author = {Thorbjørn Erik Køppen Christensen and Maja Østergaard and Olof Gutowski and Ann-Christin Dippel and Henrik Birkedal},
}

@book{Cornell2003,
    title={{The iron oxides: structure, properties, reactions, occurrences, and uses}},
    author={Cornell, Rochelle M and Schwertmann, Udo and others},
    year={2003},
    publisher = {John Wiley \& Sons, Ltd},
    doi= {10.1002/3527602097},
    isbn = {9783527602094},
    url = {https://onlinelibrary.wiley.com/doi/abs/10.1002/3527602097}
}

@article{Costa2020,
  title={{The internal structure and geodynamics of Mars inferred from a 4.2-Gyr zircon record}},
  author={Costa, Maria M and Jensen, Ninna K and Bouvier, Laura C and Connelly, James N and Mikouchi, Takashi and Horstwood, Matthew SA and Suuronen, Jussi-Petteri and Moynier, Fr{\'e}d{\'e}ric and Deng, Zhengbin and Agranier, Arnaud and others},
  journal={Proceedings of the National Academy of Sciences},
  volume={117},
  number={49},
  pages={30973--30979},
  year={2020},
  publisher={National Academy of Sciences}
}

@article{Curetti2011,
    url = {https://doi.org/10.2138/am.2011.3604},
    title = {{High-pressure structural evolution and equation of state of analbite}},
    author = {Nadia Curetti and Lindsay M. Sochalski-Kolbus and Ross J. Angel and Piera Benna and Fabrizio Nestola and Emiliano Bruno},
    pages = {383--392},
    volume = {96},
    number = {2-3},
    journal = {American Mineralogist},
    doi = {10.2138/am.2011.3604},
    year = {2011},
}

@article{Farley2020,
  title={Mars 2020 mission overview},
  author={Farley, Kenneth A and Williford, Kenneth H and Stack, Kathryn M and Bhartia, Rohit and Chen, Al and de la Torre, Manuel and Hand, Kevin and Goreva, Yulia and Herd, Christopher DK and Hueso, Ricardo and others},
  journal={Space Science Reviews},
  volume={216},
  number={8},
  pages={142},
  year={2020},
  publisher={Springer}
}

@article{Farley2022,
  title={{Aqueously altered igneous rocks sampled on the floor of Jezero crater, Mars}},
  author={Farley, KA and Stack, KM and Shuster, DL and Horgan, BHN and Hurowitz, JA and Tarnas, JD and Simon, JI and Sun, VZ and Scheller, EL and Moore, KR and others},
  journal={Science},
  volume={377},
  number={6614},
  pages={eabo2196},
  year={2022},
  publisher={American Association for the Advancement of Science}
}

@article{Fedrigo2018,
    author = {Fedrigo, A. and others},
    title = {{Investigation of a {Monturaqui} Impactite by Means of Bi-Modal X-ray and Neutron Tomography}},
    journal = {Journal of Imaging},
    volume = {4},
    year = {2018},
    number = {5},
    article-number = {72},
    url = {https://www.mdpi.com/2313-433X/4/5/72},
    issn = {2313-433X},
    doi = {10.3390/jimaging4050072}
}

@article{Feldman2002,
  title={{Global distribution of neutrons from Mars: Results from Mars Odyssey}},
  author={Feldman, WC and Boynton, WV and Tokar, RL and Prettyman, TH and Gasnault, O and Squyres, SW and Elphic, RC and Lawrence, DJ and Lawson, SL and Maurice, S and others},
  journal={Science},
  volume={297},
  number={5578},
  pages={75--78},
  year={2002},
  publisher={American Association for the Advancement of Science}
}

@article{Fujiwara2006,
  title={The rubble-pile asteroid Itokawa as observed by Hayabusa},
  author={Fujiwara, Akira and Kawaguchi, J and Yeomans, DK and Abe, M and Mukai, T and Okada, T and Saito, J and Yano, H and Yoshikawa, M and Scheeres, DJ and others},
  journal={Science},
  volume={312},
  number={5778},
  pages={1330--1334},
  year={2006},
  publisher={American Association for the Advancement of Science}
}

@article{Frølich2016,
    author = "Fr{\o}lich, S. and Leemreize, H. and Jakus, A. and Xiao, X. and Shah, R. and Birkedal, H. and Almer, J. D. and Stock, S. R.",
    title = "{Diffraction tomography and Rietveld refinement of a hydroxyapatite bone phantom}",
    journal = "Journal of Applied Crystallography",
    year = "2016",
    volume = "49",
    number = "1",
    pages = "103--109",
    doi = {10.1107/S1600576715022633},
    url = {https://doi.org/10.1107/S1600576715022633},
}

@article{Gattacceca2014,
    author = {Gattacceca, J. and Rochette, P. and Scorzelli, R. B. and Munayco, P. and Agee, C. and Quesnel, Y. and Cournède, C. and Geissman, J.},
    title = {{Martian meteorites and Martian magnetic anomalies: A new perspective from NWA 7034}},
    journal = {Geophysical Research Letters},
    volume = {41},
    number = {14},
    pages = {4859-4864},
    doi = {10.1002/2014GL060464},
    url = {https://agupubs.onlinelibrary.wiley.com/doi/abs/10.1002/2014GL060464},
    year = {2014}
}

@article{Grotzinger2015,
  title={{Deposition, exhumation, and paleoclimate of an ancient lake deposit, Gale crater, Mars}},
  author={Grotzinger, John P and Gupta, Sanjeev and Malin, Michael C and Rubin, David M and Schieber, Juergen and Siebach, Kirsten and Sumner, Dawn Y and Stack, Kathryn M and Vasavada, Ashwin R and Arvidson, Raymond E and others},
  journal={Science},
  volume={350},
  number={6257},
  pages={aac7575},
  year={2015},
  publisher={American Association for the Advancement of Science}
}

@article{Herd2025,
  title={{Sampling Mars: Geologic context and preliminary characterization of samples collected by the NASA Mars 2020 Perseverance Rover Mission}},
  author={Herd, Christopher DK and Bosak, Tanja and Hausrath, Elisabeth M and Hickman-Lewis, Keyron and Mayhew, Lisa E and Shuster, David L and Siljestr{\"o}m, Sandra and Simon, Justin I and Weiss, Benjamin P and Wadhwa, Meenakshi and others},
  journal={Proceedings of the National Academy of Sciences},
  volume={122},
  number={2},
  pages={e2404255121},
  year={2025},
  publisher={National Academy of Sciences}
}

@article{Herd2024,
  title={{The source craters of the martian meteorites: Implications for the igneous evolution of Mars}},
  author={Herd, Christopher DK and Hamilton, Jarret S and Walton, Erin L and Tornabene, Livio L and Lagain, Anthony and Benedix, Gretchen K and Sheen, Alex I and Melosh, Harry J and Johnson, Brandon C and Wiggins, Sean E and others},
  journal={Science Advances},
  volume={10},
  number={33},
  pages={eadn2378},
  year={2024},
  publisher={American Association for the Advancement of Science}
}

@article{Hewins2017,
  title={{Regolith breccia Northwest Africa 7533: Mineralogy and petrology with implications for early Mars}},
  author={Hewins, Roger H and Zanda, Brigitte and Humayun, Munir and Nemchin, Alexander and Lorand, Jean-Pierre and Pont, Sylvain and Deldicque, Damien and Bellucci, Jeremy J and Beck, Pierre and Leroux, Hugues and others},
  journal={Meteoritics \& Planetary Science},
  volume={52},
  number={1},
  pages={89--124},
  year={2017},
  publisher={Wiley Online Library}
}

@article{Hu2019,
  title={{Ancient geologic events on Mars revealed by zircons and apatites from the Martian regolith breccia NWA 7034}},
  author={Hu, Sen and Lin, Yangting and Zhang, Jianchao and Hao, Jialong and Xing, Weifan and Zhang, Ting and Yang, Wei and Changela, Hitesh},
  journal={Meteoritics \& Planetary Science},
  volume={54},
  number={4},
  pages={850--879},
  year={2019},
  publisher={Wiley Online Library}
}

@article{Humayun2013,
  title={{Origin and age of the earliest Martian crust from meteorite NWA 7533}},
  author={Humayun, M and Nemchin, Alexander and Zanda, B and Hewins, RH and Grange, Marion and Kennedy, Allen and Lorand, J-P and G{\"o}pel, C and Fieni, C and Pont, S and others},
  journal={Nature},
  volume={503},
  number={7477},
  pages={513--516},
  year={2013},
  publisher={Nature Publishing Group UK London}
}

@article{Hurowitz2025,
    author = {Hurowitz, Joel A. and others},
    title = {{Redox-driven mineral and organic associations in Jezero Crater, Mars}},
    journal =  {Nature},
    year = {2025},
    volume = {645},
    pages = {332--340},
    doi = {10.1038/s41586-025-09413-0},
    url = {https://doi.org/10.1038/s41586-025-09413-0}
}

@article{Jensen2022,
    author = "Jensen, Alexander Bernthz and Christensen, Thorbj{\o}rn Erik K{\o}ppen and Weninger, Clemens and Birkedal, Henrik",
    title = "{Very large-scale diffraction investigations enabled by a matrix-multiplication facilitated radial and azimuthal integration algorithm: {\it MatFRAIA}}",
    journal = "Journal of Synchrotron Radiation",
    year = "2022",
    volume = "29",
    number = "6",
    pages = "1420--1428",
    doi = {10.1107/S1600577522008232},
    url = {https://doi.org/10.1107/S1600577522008232},
}

@article{Jorgensen2021,
    author = {Jørgensen, J. S.  and Ametova, E.  and Burca, G.  and Fardell, G.  and Papoutsellis, E.  and Pasca, E.  and Thielemans, K.  and Turner, M.  and Warr, R.  and Lionheart, W. R. B.  and Withers, P. J. },
    title = {{Core Imaging Library - Part I: a versatile Python framework for tomographic imaging}},
    journal = {Philosophical Transactions of the Royal Society A: Mathematical, Physical and Engineering Sciences},
    volume = {379},
    number = {2204},
    pages = {20200192},
    year = {2021},
    doi = {10.1098/rsta.2020.0192},
    URL = {https://royalsocietypublishing.org/doi/abs/10.1098/rsta.2020.0192},
}

@article{Kaestner2011,
    title = {{The ICON beamline – A facility for cold neutron imaging at SINQ}},
    journal = {Nuclear Instruments and Methods in Physics Research Section A: Accelerators, Spectrometers, Detectors and Associated Equipment},
    volume = {659},
    number = {1},
    pages = {387-393},
    year = {2011},
    issn = {0168-9002},
    doi = {10.1016/j.nima.2011.08.022},
    url = {https://www.sciencedirect.com/science/article/pii/S0168900211016044},
    author = {A.P. Kaestner and S. Hartmann and G. Kühne and G. Frei and C. Grünzweig and L. Josic and F. Schmid and E.H. Lehmann},
}

@article{Kaestner2011_2,
    title = {{MuhRec—A new tomography reconstructor}},
    journal = {Nuclear Instruments and Methods in Physics Research Section A: Accelerators, Spectrometers, Detectors and Associated Equipment},
    volume = {651},
    number = {1},
    pages = {156-160},
    year = {2011},
    note = {{Proceeding of the Ninth World Conference on Neutron radiography (“The Big-5 on Neutron Radiography”)}},
    issn = {0168-9002},
    doi = {10.1016/j.nima.2011.01.129},
    url = {https://www.sciencedirect.com/science/article/pii/S0168900211002312},
    author = {Anders P. Kaestner},
}

@ARTICLE{Lagain2022,
    author = {{Lagain}, A. and {Bouley}, S. and {Zanda}, B. and {Miljkovi{\'c}}, K. and {Raj{\v{s}}i{\'c}}, A. and {Baratoux}, D. and {Payr{\'e}}, V. and {Doucet}, L.~S. and {Timms}, N.~E. and {Hewins}, R. and et al.},
    title = "{Early crustal processes revealed by the ejection site of the oldest martian meteorite}",
    journal = {Nature Communications},
    year = 2022,
    month = jul,
    volume = {13},
    pages = {3782},
    doi = {10.1038/s41467-022-31444-8},
    url = {https://ui.adsabs.harvard.edu/abs/2022NatCo..13.3782L},
}

@article{Lauretta2017,
  title={{OSIRIS-REx: sample return from asteroid (101955) Bennu}},
  author={Lauretta, DS and Balram-Knutson, SS and Beshore, Eea and Boynton, WV and Drouet d’Aubigny, C and DellaGiustina, DN and Enos, HL and Golish, DR and Hergenrother, CW and Howell, ES and others},
  journal={Space Science Reviews},
  volume={212},
  number={1},
  pages={925--984},
  year={2017},
  publisher={Springer}
}

@article{Levrard2004,
  title={{Recent ice-rich deposits formed at high latitudes on Mars by sublimation of unstable equatorial ice during low obliquity}},
  author={Levrard, Benjamin and Forget, Fran{\c{c}}ois and Montmessin, Franck and Laskar, Jacques},
  journal={Nature},
  volume={431},
  number={7012},
  pages={1072--1075},
  year={2004},
  publisher={Nature Publishing Group UK London}
}

@article{Lorand2015,
  title={{Nickeliferous pyrite tracks pervasive hydrothermal alteration in Martian regolith breccia: A study in NWA 7533}},
  author={Lorand, Jean-Pierre and Hewins, Roger H and Remusat, Laurent and Zanda, Brigitte and Pont, Sylvain and Leroux, Hugues and Marinova, Maya and Jacob, Damien and Humayun, Munir and Nemchin, Alexander and others},
  journal={Meteoritics \& Planetary Science},
  volume={50},
  number={12},
  pages={2099--2120},
  year={2015},
  publisher={Wiley Online Library}
}

@ARTICLE{Lowekamp2013,
AUTHOR={Lowekamp, Bradley C. and Chen, David T. and Ibanez, Luis  and Blezek, Daniel },
TITLE={{The Design of SimpleITK}},
JOURNAL={Frontiers in Neuroinformatics},     
VOLUME={Volume 7 - 2013},
YEAR={2013},
URL={https://www.frontiersin.org/journals/neuroinformatics/articles/10.3389/fninf.2013.00045},
DOI={10.3389/fninf.2013.00045},
ISSN={1662-5196},}

@article{Lutterotti2000,
    author = {Lutterotti, L},
    title = {{Maud: a Rietveld analysis program designed for the internet and experiment integration}},
    journal = {Acta Crystallographica Section A Foundations of Crystallography},
    year = {2000},
    doi = {10.1107/S1600576723005472},
    volume = {56},
    page = {s54}
}

@article{Mandon2023,
  title={{Reflectance of Jezero crater floor: 2. Mineralogical interpretation}},
  author={Mandon, Lucia and Quantin-Nataf, Cathy and Royer, Cl{\'e}ment and Beck, Pierre and Fouchet, Thierry and Johnson, JR and Dehouck, Erwin and Le Mou{\'e}lic, St{\'e}phane and Poulet, Fran{\c{c}}ois and Montmessin, Franck and others},
  journal={Journal of Geophysical Research: Planets},
  volume={128},
  number={7},
  pages={e2022JE007450},
  year={2023},
  publisher={Wiley Online Library}
}

@article{Martell2024,
    author = {Martell, J. and Alwmark, C. and Woracek, R. and Alwmark, S. and Hall, S. and Ferrière, L. and Daly, L. and Koch, C. Bender and Hektor, J. and Johansson, S. and Helfen, L. and Tengattini, A. and Mannes, D.},
    title = {{Combined Neutron and X-Ray Tomography—A Versatile and Non-Destructive Tool in Planetary Geosciences}},
    journal = {Journal of Geophysical Research: Planets},
    volume = {129},
    number = {2},
    pages = {e2023JE008222},
    doi = {10.1029/2023JE008222},
    url = {https://agupubs.onlinelibrary.wiley.com/doi/abs/10.1029/2023JE008222},
    year = {2024}
}

@article{Mccubbin2016,
  title={{Geologic history of Martian regolith breccia Northwest Africa 7034: Evidence for hydrothermal activity and lithologic diversity in the Martian crust}},
  author={McCubbin, Francis M and Boyce, Jeremy W and Nov{\'a}k-Szab{\'o}, T{\'\i}mea and Santos, Alison R and Tart{\`e}se, Romain and Muttik, Nele and Domokos, Gabor and Vazquez, Jorge and Keller, Lindsay P and Moser, Desmond E and others},
  journal={Journal of Geophysical Research: Planets},
  volume={121},
  number={10},
  pages={2120--2149},
  year={2016},
  publisher={Wiley Online Library}
}

@article{Mellon2004,
  title={{The presence and stability of ground ice in the southern hemisphere of Mars}},
  author={Mellon, Michael T and Feldman, William C and Prettyman, Thomas H},
  journal={Icarus},
  volume={169},
  number={2},
  pages={324--340},
  year={2004},
  publisher={Elsevier}
}

@article{Mondro2025,
  title={{Wave ripples formed in ancient, ice-free lakes in Gale crater, Mars}},
  author={Mondro, Claire A and Fedo, Christopher M and Grotzinger, John P and Lamb, Michael P and Gupta, Sanjeev and Dietrich, William E and Banham, Steven and Weitz, Catherine M and Gasda, Patrick and Edgar, Lauren A and others},
  journal={Science Advances},
  volume={11},
  number={3},
  pages={eadr0010},
  year={2025},
  publisher={American Association for the Advancement of Science}
}

@article{Morgano2018,
    author = {M. Morgano and P. Trtik and M. Meyer and E. H. Lehmann and J. Hovind and M. Strobl},
    journal = {Opt. Express},
    number = {2},
    pages = {1809--1816},
    publisher = {Optica Publishing Group},
    title = {{Unlocking high spatial resolution in neutron imaging through an add-on fibre optics taper}},
    volume = {26},
    year = {2018},
    url = {https://opg.optica.org/oe/abstract.cfm?URI=oe-26-2-1809},
    doi = {10.1364/OE.26.001809}
}

@article{Morosin1978,
    author = "Morosin, B. and Baughman, R. J. and Ginley, D. S. and Butler, M. A.",
    title = "{The influence of crystal structure on the photoresponse of iron{--}titanium oxide electrodes}",
    journal = "Journal of Applied Crystallography",
    year = "1978",
    volume = "11",
    number = "2",
    pages = "121--124",
    month = "Apr",
    doi = {10.1107/S0021889878012868},
    url = {https://doi.org/10.1107/S0021889878012868},
}

@article{Milliken2007,
  title={{Hydration state of the Martian surface as seen by Mars Express OMEGA: 2. H2O content of the surface}},
  author={Milliken, Ralph E and Mustard, John F and Poulet, Fran{\c{c}}ois and Jouglet, Denis and Bibring, Jean-Pierre and Gondet, Brigitte and Langevin, Yves},
  journal={Journal of Geophysical Research: Planets},
  volume={112},
  number={E8},
  year={2007},
  publisher={Wiley Online Library}
}

@article{Mucke1991,
    title={{The continuous alteration of ilmenite through pseudorutile to leucoxene}},
    author={M{\"u}cke, A and Chaudhuri, JN Bhadra},
    journal = {Ore Geology Reviews},
    volume = {6},
    number = {1},
    pages = {25-44},
    year = {1991},
    issn = {0169-1368},
    doi = {10.1016/0169-1368(91)90030-B},
    url = {https://www.sciencedirect.com/science/article/pii/016913689190030B},
}

@article{Muirhead2020,
  title={Mars sample return campaign concept status},
  author={Muirhead, Brian K and Nicholas, Austin K and Umland, Jeffrey and Sutherland, Orson and Vijendran, Sanjay},
  journal={Acta Astronautica},
  volume={176},
  pages={131--138},
  year={2020},
  publisher={Elsevier}
}

@article{Muttik2014,
    author = {Muttik, Nele and others},
    title = {{Inventory of H2O in the ancient Martian regolith from Northwest Africa 7034: The important role of Fe oxides}},
    journal = {Geophysical Research Letters},
    volume = {41},
    number = {23},
    pages = {8235-8244},
    doi = {10.1002/2014GL062533},
    url = {https://agupubs.onlinelibrary.wiley.com/doi/abs/10.1002/2014GL062533},
    year = {2014}
}

@BOOK{NationalAcademies2022,
  author    = "{National Academies of Sciences, Engineering, and Medicine}",
  title     = "{Origins, Worlds, and Life: A Decadal Strategy for Planetary Science and Astrobiology 2023-2032}",
  isbn      = "978-0-309-47578-5",
  doi       = "10.17226/26522",
  url       = "https://nap.nationalacademies.org/catalog/26522/origins-worlds-and-life-a-decadal-strategy-for-planetary-science",
  year      = 2023,
  publisher = "The National Academies Press",
  address   = "Washington, DC"
}

@misc{NIST_neutron,
    title = "{Neutron scattering lengths and cross sections}",
    howpublished = "\url{https://ncnr.nist.gov/resources/n-lengths/}",
    year = "1999",
    note = "Accessed: 08-08-2025"
}

@misc{NIST_xray,
    title = "{X-Ray Mass Attenuation Coefficients}",
    howpublished = "\url{https://dx.doi.org/10.18434/T4D01F}",
    year = "2004",
    note = "Accessed: 12-07-2024"
}

@article{Nyquist2016,
  title={{Rb-Sr and Sm-Nd isotopic and REE studies of igneous components in the bulk matrix domain of Martian breccia Northwest Africa 7034}},
  author={Nyquist, Laurence E and Shih, Chi-Yu and McCubbin, Francis M and Santos, Alison R and Shearer, Charles K and Peng, Zhan X and Burger, Paul V and Agee, Carl B},
  journal={Meteoritics \& Planetary Science},
  volume={51},
  number={3},
  pages={483--498},
  year={2016},
  publisher={Wiley Online Library}
}

@article{Paganin2002,
    author = {Paganin, D. and Mayo, S. C. and Gureyev, T. E. and Miller, P. R. and Wilkins, S. W.},
    title = {{Simultaneous phase and amplitude extraction from a single defocused image of a homogeneous object}},
    journal = {Journal of Microscopy},
    volume = {206},
    number = {1},
    pages = {33-40},
    doi = {10.1046/j.1365-2818.2002.01010.x},
    url = {https://onlinelibrary.wiley.com/doi/abs/10.1046/j.1365-2818.2002.01010.x},
    year = {2002}
}

@misc{Pasca2025,
    author       = {Pasca, Edoardo and others},
    title        = {{Core Imaging Library (CIL)}},
    year         = {2025},
    publisher    = {Zenodo},
    version      = {v24.3.0},
    doi          = {10.5281/zenodo.14755984},
    howpublished = {https://doi.org/10.5281/zenodo.14755984},
}

@ARTICLE{Ralph2025,
       author = {{Ralph}, Jolyon and {Von Bargen}, David and {Martynov}, Pavel and {Zhang}, Jiyin and {Que}, Xiang and {Prabhu}, Anirudh and {Morrison}, Shaunna M. and {Li}, Wenjia and {Chen}, Weilin and {Ma}, Xiaogang},
        title = "{Mindat.org: The open access mineralogy database to accelerate data-intensive geoscience research}",
      journal = {American Mineralogist},
         year = 2025,
       volume = {110},
       number = {6},
        pages = {833-844},
          doi = {10.2138/am-2024-9486},
}

@article{Santos2015,
    title = {{Petrology of igneous clasts in Northwest Africa 7034: Implications for the petrologic diversity of the martian crust}},
    journal = {Geochimica et Cosmochimica Acta},
    volume = {157},
    pages = {56-85},
    year = {2015},
    issn = {0016-7037},
    doi = {10.1016/j.gca.2015.02.023},
    url = {https://www.sciencedirect.com/science/article/pii/S0016703715001064},
    author = {A. R. Santos and Carl B. Agee and Francis M. McCubbin and Charles K. Shearer and Paul V. Burger and Romain Tartèse and Mahesh Anand}
}

@article{Schmidt2025,
  title={{Diverse and highly differentiated lava suite in Jezero crater, Mars: Constraints on intracrustal magmatism revealed by Mars 2020 PIXL}},
  author={Schmidt, Mariek E and Kizovski, Tanya V and Liu, Yang and Hernandez-Montenegro, Juan D and Tice, Michael M and Treiman, Allan H and Hurowitz, Joel A and Klevang, David A and Knight, Abigail L and Labrie, Joshua and others},
  journal={Science Advances},
  volume={11},
  number={4},
  pages={eadr2613},
  year={2025},
  publisher={American Association for the Advancement of Science}
}

@article{Simon2023,
  title={{Samples collected from the floor of Jezero Crater with the Mars 2020 Perseverance rover}},
  author={Simon, JI and Hickman-Lewis, K and Cohen, BA and Mayhew, LE and Shuster, DL and Debaille, V and Hausrath, EM and Weiss, BP and Bosak, T and Zorzano, M-P and others},
  journal={Journal of Geophysical Research: Planets},
  volume={128},
  number={6},
  pages={e2022JE007474},
  year={2023},
  publisher={Wiley Online Library}
}

@article{Stock2008,
    title = {{High energy X-ray scattering tomography applied to bone}},
    journal = {Journal of Structural Biology},
    volume = {161},
    number = {2},
    pages = {144-150},
    year = {2008},
    issn = {1047-8477},
    doi = {10.1016/j.jsb.2007.10.001},
    url = {https://www.sciencedirect.com/science/article/pii/S1047847707002535},
    author = {S.R. Stock and F. {De Carlo} and J.D. Almer},
}

@article{Tait2022,
  title={{Preliminary planning for Mars sample return (MSR) curation activities in a sample receiving facility (SRF)}},
  author={Tait, Kimberly T and McCubbin, Francis M and Smith, Caroline L and Agee, Carl B and Busemann, Henner and Cavalazzi, Barbara and Debaille, Vinciane and Hutzler, Aurore and Usui, Tomohiro and Kminek, Gerhard and others},
  journal={Astrobiology},
  volume={22},
  number={S1},
  pages={S--57},
  year={2022},
  publisher={Mary Ann Liebert, Inc., publishers 140 Huguenot Street, 3rd Floor New~…}
}

@article{Tice2022,
  title={{Alteration history of S{\'e}{\'\i}tah formation rocks inferred by PIXL x-ray fluorescence, x-ray diffraction, and multispectral imaging on Mars}},
  author={Tice, Michael M and Hurowitz, Joel A and Allwood, Abigail C and Jones, Michael WM and Orenstein, Brendan J and Davidoff, Scott and Wright, Austin P and Pedersen, David AK and Henneke, Jesper and Tosca, Nicholas J and others},
  journal={Science Advances},
  volume={8},
  number={47},
  pages={eabp9084},
  year={2022},
  publisher={American Association for the Advancement of Science}
}

@article{Valantinas2025,
  title = {{Detection of ferrihydrite in Martian red dust records ancient cold and wet conditions on Mars}},
  author={Valantinas, Adomas and Mustard, John F and Chevrier, Vincent and Mangold, Nicolas and Bishop, Janice L and Pommerol, Antoine and Beck, Pierre and Poch, Olivier and Applin, Daniel M and Cloutis, Edward A and others},
  journal={Nature Communications},
  volume={16},
  number={1},
  pages={1712},
  year={2025},
  publisher={Nature Publishing Group UK London}
}

@article{Vamvakeros2018,
    author = {Vamvakeros, A. and Jacques, S. D. M. and Di Michiel, M. and Matras, D. and Middelkoop, V. and Ismagilov, I. Z. and Matus, E. V. and Kuznetsov, V. V. and Drnec, J. and Senecal, P. and Beale, A. M.},
    title = {{5D operando tomographic diffraction imaging of a catalyst bed}},
    journal = {Nature Communications},
    year = {2018},
    doi = {10.1038/s41467-018-07046-8}, 
    url = {https://doi.org/10.1038/s41467-018-07046-8},
    volume = {9},
    pages = {4751}
}

@article{vanAarle2016,
author = {Wim van Aarle and Willem Jan Palenstijn and Jeroen Cant and Eline Janssens and Folkert Bleichrodt and Andrei Dabravolski and Jan De Beenhouwer and K. Joost Batenburg and Jan Sijbers},
journal = {Opt. Express},
number = {22},
pages = {25129--25147},
publisher = {Optica Publishing Group},
title = {{Fast and flexible X-ray tomography using the ASTRA toolbox}},
volume = {24},
month = {Oct},
year = {2016},
url = {https://opg.optica.org/oe/abstract.cfm?URI=oe-24-22-25129},
doi = {10.1364/OE.24.025129},
}

@article{Vasavada2022,
  title={{Mission overview and scientific contributions from the Mars Science Laboratory Curiosity rover after eight years of surface operations}},
  author={Vasavada, Ashwin R},
  journal={Space Science Reviews},
  volume={218},
  number={3},
  pages={14},
  year={2022},
  publisher={Springer}
}

@article{Watanabe2017,
  title={{Hayabusa2 mission overview}},
  author={Watanabe, Sei-ichiro and Tsuda, Yuichi and Yoshikawa, Makoto and Tanaka, Satoshi and Saiki, Takanao and Nakazawa, Satoru},
  journal={Space Science Reviews},
  volume={208},
  number={1},
  pages={3--16},
  year={2017},
  publisher={Springer}
}

@article{Wechsler1984,
    Author = {Wechsler, BA and Lindsley, DH and Prewitt, CT},
    Title = {{Crystal structure and cation distribution in titanomagnetites (\ce{Fe_{3-x}TixO4}) }},
    Journal = {American Mineralogist},
    Year = {1984},
    Volume = {69},
    Number = {7-8},
    Pages = {754-770},
    Publisher = {MINERALOGICAL SOC AMER},
    ISSN = {0003-004X},
}

@article{Wiens2022,
  title={{Compositionally and density stratified igneous terrain in Jezero crater, Mars}},
  author={Wiens, Roger C and Udry, Arya and Beyssac, Olivier and Quantin-Nataf, Cathy and Mangold, Nicolas and Cousin, Agn{\`e}s and Mandon, Lucia and Bosak, Tanja and Forni, Olivier and McLennan, Scott M and others},
  journal={Science advances},
  volume={8},
  number={34},
  pages={eabo3399},
  year={2022},
  publisher={American Association for the Advancement of Science}
}

@article{Wernicke2021,
  title={{Martian hydrated minerals: A significant water sink}},
  author={Wernicke, Liza J and Jakosky, Bruce M},
  journal={Journal of Geophysical Research: Planets},
  volume={126},
  number={3},
  pages={e2019JE006351},
  year={2021},
  publisher={Wiley Online Library}
}

@article{Wittmann2015,
  title={{Petrography and composition of Martian regolith breccia meteorite Northwest Africa 7475}},
  author={Wittmann, Axel and Korotev, Randy L and Jolliff, Bradley L and Irving, Anthony J and Moser, Desmond E and Barker, Ivan and Rumble III, Douglas},
  journal={Meteoritics \& Planetary Science},
  volume={50},
  number={2},
  pages={326--352},
  year={2015},
  publisher={Wiley Online Library}
}

@article{Oestergaard2023,
    author = "{\O}stergaard, Maja and Naver, Estrid Buhl and Kaestner, Anders and Willendrup, Peter K. and Br{\"{u}}el, Annemarie and S{\o}rensen, Henning Osholm and Thomsen, Jesper Skovhus and Schmidt, S{\o}ren and Poulsen, Henning Friis and Theil Kuhn, Luise and Birkedal, Henrik",
    title = "{Polychromatic neutron phase-contrast imaging of weakly absorbing samples enabled by phase retrieval}",
    journal = "Journal of Applied Crystallography",
    year = "2023",
    volume = "56",
    number = "3",
    pages = "673--682",
    month = "Jun",
    doi = {10.1107/S1600576723003011},
    url = {https://doi.org/10.1107/S1600576723003011},
}

\makeatletter
\renewcommand \thesection{S\@arabic\c@section}
\renewcommand\thetable{S\@arabic\c@table}
\renewcommand \thefigure{S\@arabic\c@figure}
\makeatother

\section{Supplementary}

\section{S1 Rietveld refinement}
Two examples of Rietveld refinement performed in Maud with a flat background. Negative values from reconstruction artefacts were set to zero.  
\begin{figure}[h]
    \centering
    \includegraphics[width=0.8\linewidth]{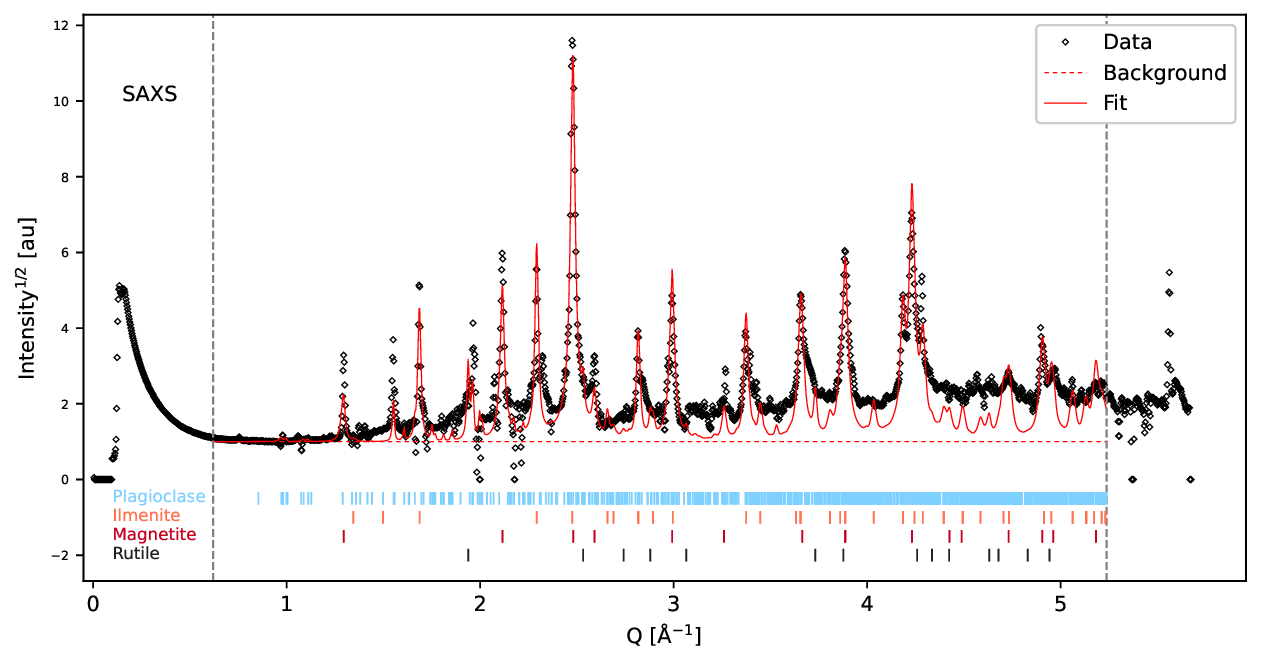}
    \caption{Rietveld refinement of H-rich phases in clast 1 performed in Maud. Vertical dashed line indicate Q-range used for Rietveld refinement. Q-range used to determine SAXS signal labelled.}
    \label{Fig.suppl_clast1}
\end{figure}

\begin{figure}[h]
    \centering
    \includegraphics[width=0.8\linewidth]{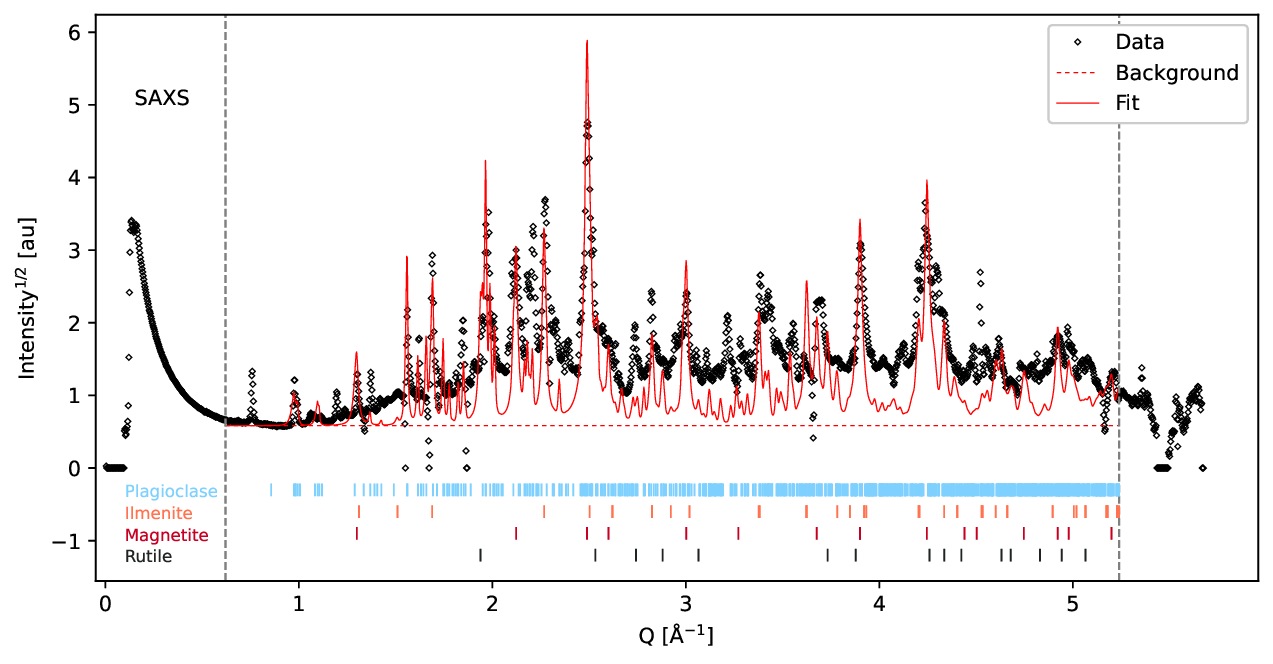}
    \caption{Rietveld refinement of H-rich phases in clast 3 performed in Maud. Vertical dashed line indicate Q-range used for Rietveld refinement. Q-range used to determine SAXS signal labelled.}
    \label{Fig.suppl_clast3}
\end{figure}

\section{S2 Calculating attenuation values}
Linear attenuation, $\mu_\text{min}$, for the minerals were calculated using 
\begin{gather}
    \mu_\text{min} = \left( \sum_i w_i \frac{\mu_i}{\rho_i} \right)\rho_\text{min} = \left( \sum_i w_i \frac{\sigma_{\text{tot},i}}{uA_i} \right)\rho_\text{min} ,
\end{gather}
where $\frac{\mu}{\rho}$ is the mass attenuation, $w_i$ are the mass percentages for the individual elements, $\sigma_{\text{tot},i}$ is the total interaction cross section of the element, $u$ is the atomic mass unit, $A_i$ is the relative atomic mass of the element, and $\rho$ is the material density of the mineral. Mass percentages are calculated based on chemical formula as seen in Tab. \ref{Tab. suppl_att_min}.
To calculate X-ray linear attenuation the weighted sum of mass attenuation values was used, at an X-ray energy of 73 keV corresponding to the mean energy of the X-ray beam used for the X-ray CT. To calculate neutron linear attenuation the weighted sum of total cross section was used at an energy of 25 meV, corresponding to the peak energy of the neutron beam used for neutron CT. The total cross section is the sum of the scattering and absorption cross sections. 

The resulting values calculated for the attenuation are shown in Tab. \ref{Tab. suppl_att_min}.
\begin{table}[h]
    \centering
    \caption{Calculated linear attenuation for the different minerals given the specified chemical compositions.}
    \begin{tabular}{l c c c}
        \hline
         Name & Chemical & X-ray att. & Neutron att. \\ 
         & composition & [cm$^{-1}$] & [cm$^{-1}$]  \\\hline
         Plagioclase, albite       & \ce{NaAl2Si2O8} & 0.5272 & 0.277  \\
         Plagioclase, anorthite    & \ce{CaAl2Si2O8} & 0.6468 & 0.270  \\
         Orthopyroxene, enstatite  & \ce{Mg2Si2O4} & 0.7410 & 0.369  \\
         Clinopyroxene, diopside   & \ce{CaMgSi2O6} & 0.8089 & 0.341  \\
         Clinopyroxene, CaFeAl     & \ce{CaFeAl2O6} & 1.199 & 0.383  \\
         Perthitic feldspar        & \ce{Na1/2K1/2AlSi3O8} & 0.564 & 0.271  \\
         Chlorapatite              & \ce{Ca5P3O12Cl} & 0.949 & 0.341  \\
         Ilmenite                  & \ce{FeTiO3} & 2.270 & 0.699  \\
         Magnetite                 & \ce{Fe3O4} & 3.010 & 0.801  \\
         Maghemite                 & \ce{Fe2O3} & 2.988 & 0.815  \\
         \hline
    \end{tabular}
    \label{Tab. suppl_att_min}
\end{table}

Finding theoretical attenuation values for the clasts and the matrix was done using weighted means of the linear attenuation $\mu_\text{min}$ for each mineral. The matrix composition was taken to be 38\% plagioclase feldspar, 25.4\% pyroxene, 18.2\% clinopyroxenes, 9.7\% iron oxides, 4.9\% alkali feldspars, and 3.7\% apatite. The resulting values are shown in Tab. \ref{Tab. suppl_att_clast}. 
\begin{table}[h]
    \centering
    \caption{Measured and calculated linear attenuation values for clasts 1, 2, and 3 as well as an area of matrix without highly attenuating phases.}
    \begin{tabular}{lcccc}
    \hline
     & \multicolumn{2}{l}{\textbf{Measured}} & \multicolumn{2}{l}{\textbf{Theory}} \\
     & \multicolumn{1}{l}{no H {[}cm$^{-1}${]}} & \multicolumn{1}{l}{H {[}cm$^{-1}${]}} & \multicolumn{1}{l}{no H {[}cm$^{-1}${]}} & \multicolumn{1}{l}{H {[}cm$^{-1}${]}} \\ \hline
    \textbf{X-ray} & & & & \\
    Clast 1          & 2.1    & 3.5    & 2.1    & 2.7    \\
    Clast 2          & 1.8    & 3.4    & 1.9    & 1.7    \\
    Clast 3          & 1.4    & 3.4    & 1.3    & 1.7    \\
    Clast average    & 1.8    & 3.4    & 1.7    & 2.0    \\
    Matrix           & 0.96   & -      & 0.96   & -       \\ \hline
    \textbf{Neutron} & & & &       \\
    Clast 1          & 0.56    & 1.9    & 0.64   & 0.76    \\
    Clast 2          & 0.57    & 1.7    & 0.58   & 0.53    \\
    Clast 3          & 0.58    & 2.0    & 0.46   & 0.54     \\
    Clast average    & 0.57    & 1.9    & 0.56   & 0.61    \\
    Matrix           & 0.45    & -      & 0.37   & -   \\ \hline              
    \end{tabular}
    \label{Tab. suppl_att_clast}
\end{table}

When estimating the OH content of the clasts and matrix, mass attenuation values for the minerals were used, and the resulting linear attenuation estimated using the density of the meteorite.

\begin{equation}
    \mu_{\text{clast}} =  \left( \sum_\text{min} w_\text{min} \frac{\mu_\text{min}}{\rho_i} \right)\rho_\text{meteorite}
\end{equation}

\end{document}